\documentclass{emulateapj}

\def\be{\begin{equation}}
\def\ee{\end{equation}}

\begin{document}

\title{A Method for Mapping the Temperature Profile of X-ray
Clusters through Radio Observations}

\author{Gilbert P. Holder and Abraham Loeb\altaffilmark{1}}
\affil{Institute for Advanced Study, Einstein Drive, Princeton NJ 0854 085400}

\submitted{to be submitted to ApJL}
\email{holder@ias.edu}
\altaffiltext{1}{Guggenheim Fellow; on sabbatical leave from the
Astronomy Department, Harvard University.}

\begin{abstract}
Many of the most luminous extragalactic radio sources are located at the
centers of X-ray clusters, and so their radiation must be scattered by
the surrounding hot gas.  We show that radio observations of the
highly-polarized scattered radiation (which depends on the electron
density distribution) in combination with the thermal
Sunyaev-Zeldovich effect (which measures the electron pressure
distribution), can be used to determine the radial profile of the
electron temperature within the host cluster. The sensitivity levels
expected from current instruments will allow radio measurements of 
mass-weighted cluster temperature profiles to better than 
a $\sim 1$ keV accuracy,
as long as the central radio source is steady over several million
years.  Variable or beamed sources will leave observable signatures
in the scattered emission. 
For clusters with a central point source brighter than $\sim
1$ mJy, the scattered polarization signal is stronger than competing
effects due to the cosmic microwave background.
\end{abstract}

\section{Introduction}

Massive galaxy clusters often host a bright radio source at their
center (e.g., Zhao, Burns \& Owen 1989). \nocite{zhao89} 
A small fraction of the radiation emitted by the
central source is expected to scatter off the surrounding intracluster
electrons and obtain a high level of polarization. The radial profile
of the scattered radiation depends only on the electron density
distribution within the cluster, in difference from the thermal
Sunyaev-Zeldovich (SZ) effect which depends on the electron pressure
distribution (see Carlstrom, Holder \& Reese 2002 for a recent
review). \nocite{carlstrom02} In this {\it Letter}, we show that the
temperature profile of the intracluster gas can therefore be
determined by means of radio observations alone, through the combined
detection of the polarized scattered light from a central source and
the SZ effect.  This method for inferring cluster temperatures is
particularly powerful for distant clusters, where the alternative
option of using an X-ray telescope requires very long integration
times and is less practical.

The typical optical depth for Thomson scattering through the core of
massive X-ray clusters is $\sim 0.01$ \citep{mason00}, and so roughly a percent
of the observed flux from the central radio source should appear in
the form of a diffuse radio halo around it.  To give a concrete
example, a point source observed to have a flux density of 1 Jy 
in a cluster at a redshift $\sim 0.2$
would generate a $\sim 10$ mJy radio halo on an angular scale of $\sim
1^\prime$, assuming it is emitting isotropically.
Despite its faintness, the diffuse radio halo could be separated from
contaminating sources due to the high amplitude ($\sim 50\%$) and the
large-scale tangential alignment of its polarization 
pattern \citep{sunyaev80b,sunyaev82,sazonov99}.
The typical intrinsic polarization of individual point
sources is only a few percent, making the central point source a much
less significant contaminant in the polarization map than it is in the 
intensity map. With current instruments such as the VLA achieving
stability over a dynamic range of $10^5$ \citep{perley99}, the halo
signal should be detectable.

Several groups are currently constructing radio telescopes optimized to
search for the SZ signature of clusters of galaxies (
e.g., Carlstrom, Holder \& Reese 2003) \nocite{carlstrom02}.  The
planned SZ surveys will be efficient at finding clusters of galaxies
up to relatively high redshifts ($z \sim 1$), from which the received
X-ray flux would be weak.  In addition to the reduction in the X-ray
flux due to the large cosmological distance, the prominence of galactic
outflows (due to star formation and quasar activity) at $z\sim 1-2$
will likely disperse somewhat the intracluster gas and make the cluster less
X-ray luminous\citep{evrard91,kaiser91}.
These problems are not nearly as acute for SZ observations \citep{holder01},
raising the possibility of strong SZ sources that are difficult to follow 
up with X-ray telescopes.
Our proposed method provides an important alternative route to probing
the density and temperature distributions of the intracluster gas in
the most massive clusters that will be discovered by future SZ
surveys.

Most SZ-detected clusters will probably not have a suitable central point 
source for obtaining detailed images of the reflected images.
In nearby clusters, roughly one quarter of clusters host a radio source with
radio power greater than $10^{24.5}$ W/Hz at a wavelength of 21 cm
\citep{ledlow95}. At $z \sim 0.5$ this corresponds to roughly 10 mJy, so it
might be expected that the polarized emission might be detectable in a sizable
fraction. However, the observed radio luminosity function
\citep{ledlow96} at high power falls off quickly (dN/dlnP$\propto P^{-1.5}$),
so the fraction of sources with high enough radio powers to give a strong signal
is likely a few percent, depending on how much the cluster radio source
luminosity function evolves with redshift. From recent observations
at 30 GHz (see Bennett et al. 2003 for a recent compilation) of radio
point sources, one would expect roughly 300 sources brighter than 0.1 Jy
per steradian. Deep SZ surveys expect roughly 30000 sources per steradian
\citep{carlstrom02}, so if every bright radio source is embedded in a cluster 
the reflected emission should be significant for roughly 1\% of detected 
SZ clusters. 

The projected profile of the scattered radio emission is a direct
probe of the electron density run with radius $n_e(r)$. By comparison,
the X-ray emissivity is sensitive mainly to $n_e^2$, while the SZ
effect probes $n_e T$, where $T(r)$ is the radial profile of the
electron temperature. The combination of SZ measurements with radio
polarization measurements allows a direct determination of the
electron temperature profile, while the combination of X-ray
observations with polarization halo measurements allows a direct
distance determination, based on the same method that is currently
used with X-ray and SZ observations \citep{reese02}. Finally, the combination
of X-ray, SZ, and polarization observations would allow a consistency
check on the underlying simplifying assumptions, such as the spherical
geometry of the cluster or the steadiness of the central source.

Our proposed method could be complicated by temporal variability of
the radio source, because of the geometric time delay that the
scattered light acquires.  Images of extended radio jets indicate that
extended radio sources are often stable over those timescales 
\citep{begelman84}. However, if central cluster sources do
vary on short timescales, then an independent
mapping of the electron distribution around these sources (e.g. based
on deep X-ray observations) could be used to read off their temporal
variability as a function of age from the brightness distribution as a
function of projected radius in their scattered radio surface brightness.  
If the source is strongly beamed in a direction not aligned with our 
line of sight, it is possible to use the reflected emission as a probe 
of both the source luminosity and the surrounding medium
(see, for example Chiaberge et al. 2003 for an application using
X-ray and ultraviolet imaging).  
However, for simplicity, we focus in this publication 
on the subject of steady sources emitting isotropically. 

In \S 2 we calculate the polarization of the scattered radio halo,
while \S\ 3 outlines a synthetic observational procedure to test the
prospects for temperature measurements by the above method.  We close
with a discussion of some observational and theoretical hurdles which
must be overcome as well as a summary of the many potentially
important applications of measuring diffuse radio polarization in
galaxy clusters.

\section{Scattered Radio Halo Around a Central Point Source}

We start with the idealized case of a steady point source of observed
spectral flux density $S_\nu$ and spectral luminosity $L_\nu$ (per observed
frequency $\nu$) at the center of a spherical cluster.
We define a coordinate system centered on the cluster with the
$z$-axis along the line of sight and calculate the circularly
symmetric brightness profile of the scattered radiation. An
infinitesimal cell of physical area $dx dy$ on the sky contains
$dN_e=n_e dx dy dz$ electrons and therefore has a physical effective
cross-section to incoming photons of $dN_e \sigma_T$, where $\sigma_T$
is the Thomson cross section. The total spectral luminosity that will
therefore be scattered by this cell is $dN_e (\sigma_T/4 \pi r^2)
L_\nu$, where $r$ is the 3D radius from the cluster center. The
radiation scattered along the $z$-axis toward the observer will be
preferentially polarized in the direction perpendicular (on the sky)
to the cylindrical radius vector. This is particularly easy to see for
a parcel of electrons located in the plane of the sky but offset from
the central source. An initially unpolarized source will have equal
components of the electric field ${\bf E}$ oriented in the plane of
the sky (perpendicular to the radius vector) and along the $z$-axis.
Scattering of the $z$-polarized photons into our line of sight will be
suppressed, leading to the scattered radiation being completely
polarized in the direction perpendicular to the radius vector in the
plane of the sky. In general, the suppression factor of the component
along the cylindrical radius vector is proportional to $\cos^2\theta$,
where $\theta$ is the angle between the spherical radius vector and
the $z$-axis.

Defining an azimuthal angle on the sky $\phi$ and a coordinate system
for the Stokes parameter $Q$ (in units of intensity, not flux) such
that the maximum of $Q$ is obtained at $\phi=0$, we find 
(see also Sazonov \& Sunyaev 1999, Sunyaev 1983)
\nocite{sazonov99,sunyaev82} that for a
general distribution of electron number density $n_e(\vec{r})$
\begin{equation}
%\begin{eqnarray}
Q(R,\phi) = {3 \sigma _{T} \over 16 \pi} \left({S_\nu \over {\rm
Jy}}\right) \cos{2\phi} \int_{-\infty}^{+\infty} dz {d_A^2 R^2 \over
r^4 } n_e(R,\phi,z) \quad {\rm {Jy\over sr}} \quad ,
\label{eqn:q_gen}
\end{equation}
where $R$ is the cylindrical physical (not angular) radius in the
plane of the sky, $r= \sqrt{R^2 + z^2}$, and $d_A(z)$ is the angular
diameter distance to the source (arising here from the conversion of
physical area on the sky, $dx dy$, to solid angle).

For an electron number density profile following the
``$\beta$-profile'', where $n_e = n_{e0} (1+r^2/r_c^2)^{3\beta/2}$
\citep{cavaliere76}, one obtains for the Stokes $Q$ parameter [see,
Gradshteyn \& Ryzhik 1980, 3.259(3), p. 299]
\nocite{gradshteyn80}
%\begin{equation}
\begin{eqnarray}
Q(R,\phi) =& {3 \sigma _{T} \over 16 \pi} \left({S_\nu\over {\rm
Jy}}\right) \cos{2\phi} {d_A^2 \over R^2} {n_{e0} r_c \over (1 +
R^2/r_c^2)^{(3\beta-1)/2} } \\ 
\nonumber   & \times 
	_2F_1[2,{1\over 2};{3 \over 2} \beta +2;1- {r_c^2+R^2 \over R^2}] \\
\nonumber & \times B[{1 \over 2}, {3\over 2}(1+\beta)] 
\quad 
{\rm {Jy\over sr}},
\label{eqn:q_beta}
%\end{equation}
\end{eqnarray}
where $B(x,y)$ is the Beta function and $_2 F_1(a,b;c;z)$ is the
hypergeometric function. A similar expression in terms of
a Beta function and $_2 F_1$ can be obtained for the total scattered
intensity. For the case of $\beta=3/2$ an explicit expression can
be found in Sunyaev (1983). Note that the source redshift does not
enter explicitly into this expression, since the intrinsic source luminosity
and reflected luminosity share the same redshift factors.

Typical parameter values are $d_A \sim 1
\,h^{-1}\,$Gpc for a source at a cosmological distance, and $n_e \sim
0.01~{\rm cm^{-3}}$, $r_c \sim 150\,h^{-1}\,$kpc for typical cluster
cores.
For a 1 Jy source, these parameters correspond to roughly 1 mJy/${\rm
arcmin}^2$ at about a core radius ($\sim 0.5^\prime$) from the central
source.
At an observing frequency of $\nu=30$ GHz, the relevant conversion to
brightness temperature is 1 Jy/${\rm arcmin}^2$ = 0.4 K, and so the
scattered intensity is equivalent to nearly a mK polarized temperature
signal. In a signal-to-noise sense, this is an easy signal to detect
for the current generation of instruments. The difficulty is that such
a signal arises from a field that by construction has a 1 Jy source at
the center which must be stably removed. Typically, the 1 Jy source
may also intrinsically produce tens of mJy of polarized flux. Note
that the polarized scattered radiation will be more centrally concentrated
than the projected gas distribution due to the $1/r^2$ fall off of the 
source flux as seen by the cluster electrons.

\section{Prospects for Temperature Measurements}

We examine the feasibility of temperature measurements by constructing
synthetic observations of galaxy clusters with steady point sources
embedded at the center and emitting isotropically. 
We first use $\beta$-models and then examine
clusters in three-dimensional cosmological simulations with
hydrodynamics. Throughout our discussion, we consider a 1 Jy source
observed at 30 GHz. All properties of the polarized radio halo, scale
directly with the point source flux. The scattered intensity from the
point source may exceed the SZ signal amplitude in the core of the
cluster.  In principle, high-frequency SZ observations are not
necessary, since the scattered intensity can be reconstructed from the
polarization and removed from the SZ flux.

First we assume a $\beta$-profile for the radial profiles of the gas
temperature and density, and construct the brightness profiles for the
SZ effect and polarized radio halo around the central source. Rather
than tailor our predictions closely to a particular instrument, we
assume that instruments such as 
SZA\footnote{http://astro.uchicago.edu/SZE/survey.html},
CARMA\footnote{http://www.mmarray.org},
AMI\footnote{http://www.mrao.cam.ac.uk/telescopes/ami},
Amiba\footnote{http://www.asiaa.sinica.edu.tw/amiba},
EVLA\footnote{http://www.aoc.nrao.edu/evla},
or ALMA\footnote{http://www.alma.nrao.edu/}
may generically be thought of having sensitivity to a range of angular
scales from roughly arcsecond scales up to a few arcminutes with up to
${\rm \mu K}$ sensitivity. Note that SZA and AMI are not currently planned
to be used for polarization studies, but it would not be difficult to adapt
these instruments for polarization studies. 

The details will depend crucially on the observing frequency.  
In general, large scale emission will be filtered out. We assume that
this filtering is on a scale of ten arcminutes or a radius in the
$u$--$v$ plane corresponding to $\sim 300$ wavenumbers (in units of the
observing wavelength)
and that there is no sensitivity to larger scales (i.e., we assume a
high pass filter). At cm to mm wavelengths this is effectively
required to avoid contamination from primary cosmic microwave
background (CMB) anisotropies. In addition, high angular resolution is
required to isolate the central point source.  We assume a maximum
$u$--$v$ radius of 3000 wavenumbers and an equal noise contribution
per $log$ interval in $u-v$ radius between the above inner and outer cutoffs. 
Noise was added such that the resulting noise in each map was 1 ${\rm \mu K}$ 
and beam effects were ignored. An observing frequency in the Rayleigh--Jeans
was assumed for the SZ component.

We assumed a $\beta$-profile for each of the electron number density
profile and the pressure profile, with a central electron number density of
0.01 ${\rm cm}^{-3}$, a central temperature of $T_{e0}=5$ keV, a
common core radius of $0.75^\prime$ for both $n_e(r)$ and $T(r)$,
$\beta_e=2/3$ for the electron density profile and $\beta_P=1$ for
the pressure profile. The cluster was assumed to be at $z=0.5$.
A realization of ``data'' was generated and then fit to a model with two
$\beta$-models, with the centers of the profiles constrained to be at
the cluster center. Each $\beta$-model had three parameters, and the
presence of a point source at the center was allowed in terms of two
free parameters (flux and polarized flux). A Markov chain Monte-Carlo
method was used to estimate constraints on the temperature profile as
a function of radius. Specifically, random steps were taken in the 8-D
parameter space and accepted if the proposed point was higher
probability or rejected according to a probability set by how less
likely the proposed point was than the current point. After many
samples, the list of points in the chain provides a sampling of the
likelihood function. For each point in the chain the temperature
profile $T(r)$ was constructed. The resulting distribution of
temperature profiles is shown in Figure 1, where all parameters have
effectively been marginalized over.
There is a dramatic reduction in the uncertainties if the central
point source flux is assumed to be known both in terms of polarized
emission and the SZ signal; this simply reflects the fact that the
polarized intensity is highly concentrated, and therefore can be
partially mimicked by a point source. Given the importance of the
central flux it seems prudent to simply marginalize over the unknown
flux and accept that the temperature uncertainties will be on the
order of $\sim 1$ keV, comparable to X-ray spectroscopic temperature
uncertainties.

\begin{figure}
\plotone{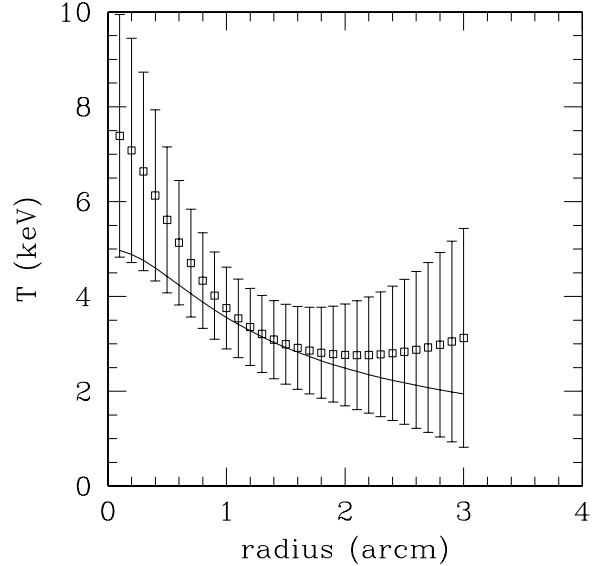}
\caption{Reconstructed temperature profile for a single realization
(assuming the electron density and temperature profiles are both
$\beta$-models) with approximate 1-$\sigma$ uncertainties. Neighboring
points are highly correlated. The true input profile is shown as a
solid line.  }
\end{figure}

Real clusters are unlikely to follow ideal spherically-symmetric
$\beta$-profiles, and so we repeated the calculation for clusters
extracted from numerical simulations produced by the Hydra 
Consortium\footnote{http://hydra.susx.ac.uk} and made publicly available 
by the Virgo Consortium\footnote{http://virgo.sussex.ac.uk/clusdata.html} 
on the web.  The simulation data shows
deviations from the ideal $\beta$-profiles that are expected in
nature.  These simulations have been used \citep{dasilva99} to study the
statistics of SZ maps.
Maps of two clusters, both in SZ and polarized emission, are shown in
Figure 2. We repeated the above calculation, generating synthetic data,
fitting to two $\beta$-profiles and comparing to the true temperature
profile. Results were qualitatively similar to those for the case of the 
spherical $\beta$ clusters. 
In the simulated clusters the mass-weighted temperature within
three arcminutes was recovered with statistical errors of roughly 1 keV.

\begin{figure}
\plottwo{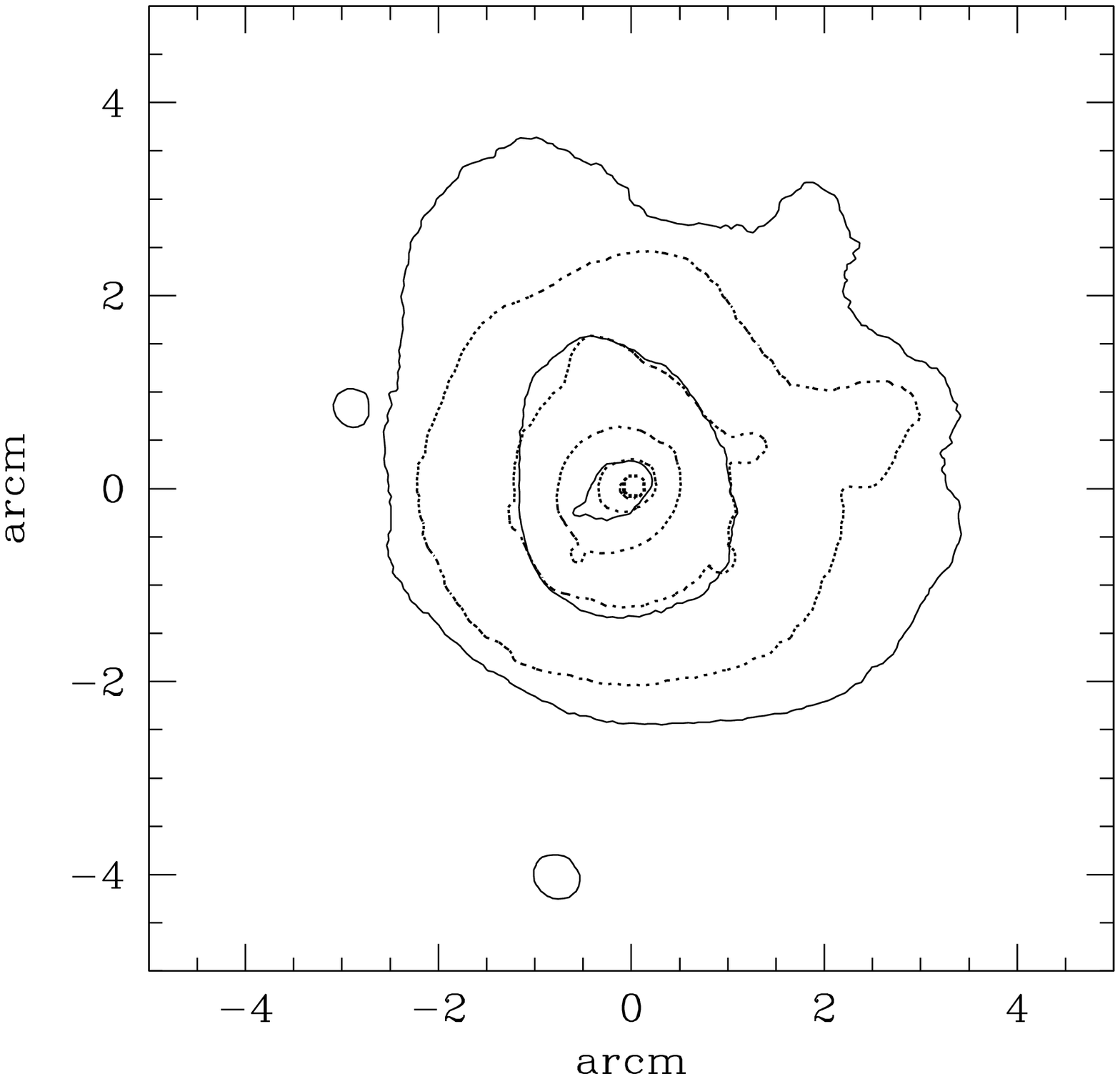}{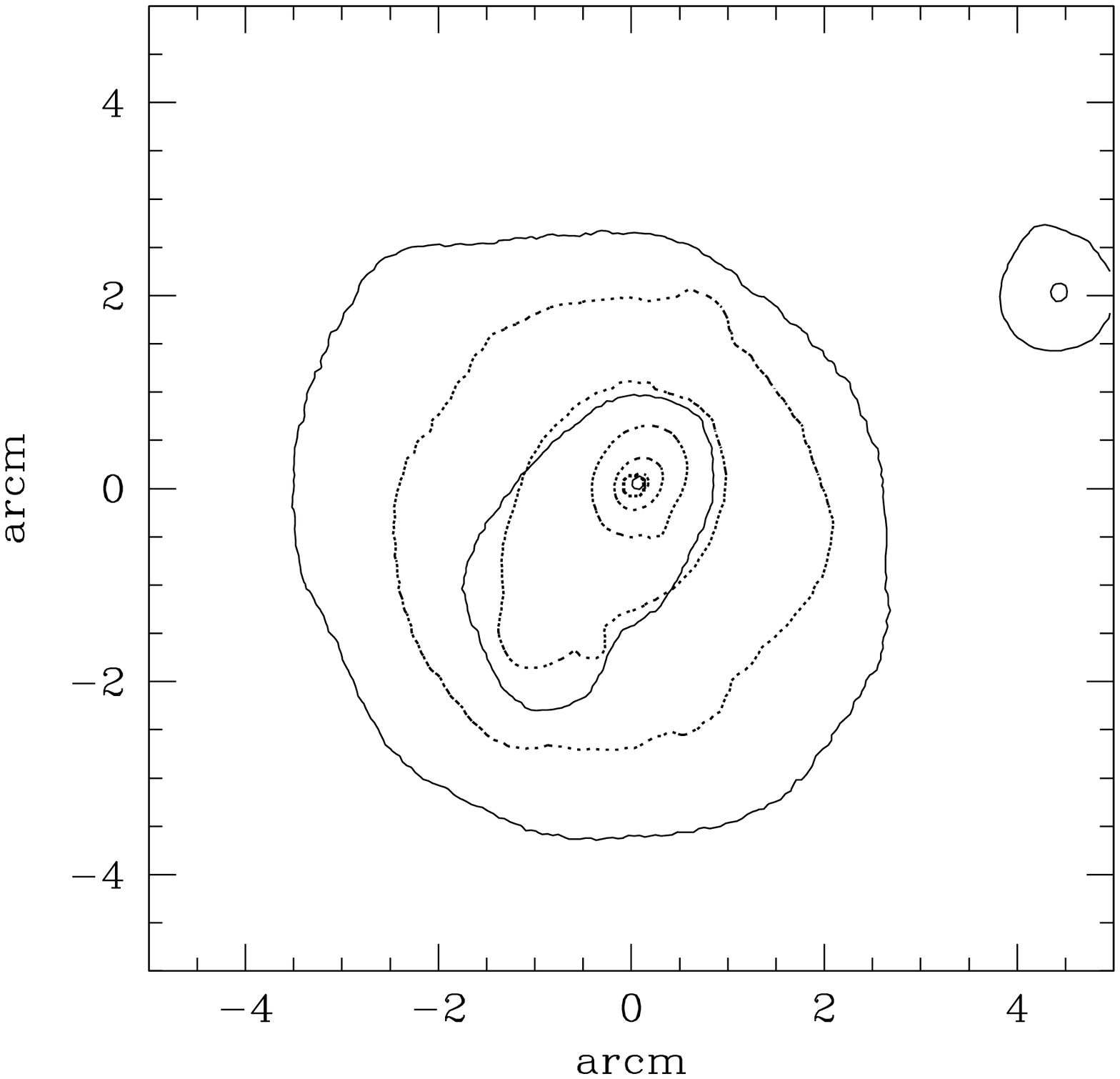}
\caption{ Maps of thermal SZ (solid) and polarized intensity (dotted)
for the most massive (left) and second most massive (right) clusters
in the simulation archive. Outermost contours are 1 ${\rm \mu K}$ at 30 GHz
and they increase in factors of 10 inward. The polarized intensity
assumes a central 1 Jy source, and that the conversion to temperature
units is 1 ${\rm \mu K}$ for 0.43 ${\rm \mu Jy}$ per square arcminute.
The direction of polarization at each point is tangential to the cluster
center.  }
\end{figure}

We have not attempted a detailed search to optimize the spatial
sensitivity or frequency coverage. Contaminating and confusing sources
will be less or more important at different spatial scales or
observing frequencies. At low frequencies there could be confusion
with diffuse synchrotron emission by the cluster or extended jet
structures, while higher frequencies will run into both a relative
dearth of very bright central point sources and dust emission from our
Galaxy as well as distant star-forming galaxies \citep{fischer93}. 

\section{Discussion}

We have shown that radio measurements alone can be used to reliably
measure temperature profiles of the gas in X-ray clusters. The current
generation of radio instruments is already sufficient in terms of
sensitivity, but the presence of a 1 Jy source in the center of the
field will no doubt make sub-mJy polarized studies challenging. With
the VLA performing at dynamic ranges exceeding $10^5$
\citep{perley99}, it is clear that the presence of the central source
does not pose an inherent limitation to our method.

With typical scattered fluxes of $\sim 1\%$ of the central source
flux, it can be expected that the polarized scattered emission will
exceed the ${\rm \mu K}$ level for all central sources brighter than about 1
mJy at 30 GHz. This is brighter than the {\em peak} emission due to
the cluster scattering of the primary CMB quadrupole, which is itself
significantly larger than any anisotropies due to thermal or kinetic
SZ effects \citep{sazonov99}.  A central point source is by far the
single largest source of a quadrupole anisotropy in the radiation
field for many clusters at radio frequencies. 
Therefore, detection of the scattered radio halo could provide
an intermediate target for an attempt to measure these smaller
signals. To the extent that the radio spectrum is constant over the
age of the source, the spectrum of the polarized emission should be
the same as that of the central source, providing an important
consistency check.

The nearly circular polarization pattern should provide a simple
template for improving the detectability of the signal, and the
compact angular scale implies that the CMB anisotropy is not a serious
contaminant.  Many clusters show signs of radio halos or radio relics
\citep{giovannini02}.  These generally have steep spectra, so it may
be difficult to observe the polarized emission reflected from the
central source at wavelengths longer than a few cm. Jets associated
with the central source may be significant in the inner arcseconds.
The largest contaminant will likely be the systematic problem of
removing the very bright central source sufficiently accurately and
stably so that it would not affect the measurements in the entire
image.

Nearby clusters are too large (in angular size) to have a significant surface
brightness so the studies considered here are probably
most suitable for clusters at redshifts $z\gtrsim 0.2$. 
For studies of the SZ effect it should be noted that all
sources above $\sim 1$ {\rm mJy} will contaminate the immediate vicinity of
the point source at ${\rm \mu K}$ levels and therefore point source removal
must be handled carefully. If bright radio sources are significantly
beamed, then some radio SZ observations could be contaminated by this
reflected emission even for sources that do not appear to be very
luminous.

While we have outlined the potential for measuring temperature
profiles, an equally exciting application would be to measure
the time evolution of the radio flux from the central source. With
cluster radii of roughly 1 Mpc the light travel times are a few million 
years.  With ${\rm \mu K}$ sensitivities to polarized emission it will be
possible to measure the age of the central source by simply finding an
edge to the polarized emission; with independent temperature
measurements, either from SZ spectral studies or X-ray spectroscopy,
it would be possible to measure the time evolution of the central
radio source. Surfaces of scattering
corresponding to constant central source emission time are paraboloids
\citep{sazonov03}, and so a good
mapping of the three-dimensional electron distribution will be
necessary.

In summary, the extremely high sensitivity and stability of current
and near-future instruments should enable the use of the intracluster
medium as a reflection nebula around bright point sources. This should
provide some insight about the geometry of merging subclumps and a
measure of the cluster electron density distribution in addition to
the important applications outlined above. The observations present
some unique challenges, but the potential scientific rewards are
compelling.

\acknowledgements{G.P.H. is supported by a W.M. Keck Fellowship.
A.L. acknowledges support from the Institute for Advanced Study at
Princeton and the John Simon Guggenheim Memorial Fellowship.  This
work was also supported in part by NSF grants AST-0071019, AST-0204514
and NASA grant NAG-13292 (for A.L.).  We thank the Hydra and Virgo
consortia for making their simulation products publicly available in
an accessible format.   }

\end{document}